# Elasticity-associated rebinding rate of molecular bonds between soft elastic media

Qiangzeng Huang, Jizeng Wang[1]

*Key Laboratory of Mechanics on Disaster and Environment in Western China, Ministry of Education, College of Civil Engineering and Mechanics, Lanzhou University, Lanzhou, Gansu 730000, China*

**Abstract**

A quantitative understanding of how cells interact with their extracellular matrix via molecular bonds is fundamental for many important processes in cell biology and engineering. In these interactions, the deformability of cells and matrix are usually comparable with that of the bonds, making their rebinding events globally coupled with the deformation states of whole systems. Unfortunately, this important principle is not realized or adopted in most conventional theoretical models for analyzing cellular adhesions. In this study, we considered a new theoretical model of a cluster of ligand–receptor bonds between two soft elastic bodies, in which the rebinding rates of ligands to receptors are described, for the first time, by considering the deformation of the overall system under the influence of bond distributions. On the basis of theory of continuum mechanics and statistical mechanics, we obtained an elasticity-associated rebinding rate of open bonds in a closed analytical form that highly depends on the binding states and distributions of all other bonds, as well as on the overall deformation energy stored in the elastic bodies and all closed bonds. On the basis of this elasticity-associated rebinding rate and by performing Monte Carlo simulations, we uncovered new mechanisms underlying the adhesion stability of molecular bond clusters associated with deformable elastic bodies. Moreover, we revealed that the rebinding processes of molecular bonds is not only dependent on interfacial separation but is related to overall energy. This newly proposed rebinding rate may substantially improve our understanding of how cells adapt to their microenvironments by adjusting their mechanical properties through cytoskeleton remodeling and of how we can accurately calibrate the measurements of adhesion strength of molecular bond clusters between soft media.

---
[1] Corresponding author. *E-mail address:* jzwang@lzu.edu.cn (J. Wang).

## 1. Introduction

For most anchorage-dependent cells, attachment to cells or extracellular matrix (ECM) through the formation of focal adhesions (FAs) via clusters of transmembrane ligand–receptor bonds is essential for normal physiological processes, including growth, proliferation, mitosis, migration, and differentiation (Martino et al., 2018; Pelham and Wang, 1997; Lo et al., 2000; Discher et al., 2005; Razinia, et al., 2017). In the past decades, researchers have exerted immense efforts to quantitatively understand and experimentally measure the responses and behaviors of cells interacting with ECM.

Researchers have experimentally demonstrated that stiff matrices favor cells by a large spreading area, a strong FA, and a low fluctuation along the cell periphery (Pelham and Wang, 1997), consistent with the acknowledged ability of cells to migrate toward a stiffer area as grown in a varying stiffness matrix (Lo et al., 2000). Alternatively, a general consensus exists that tissue-specific progenitors, such as neural stem cells, pre-osteoblasts, myoblasts, and adult cardiac progenitor, acquire the given phenotype only when cultured on matrices that resemble the physiological and characteristic stiffness of the tissue they belong to (Saha, et al., 2008; Justin and Engler, 2011; Engler, et al., 2004; Mosqueira, et al., 2014). These two phenomena contradict each other as the former asserts that stiff matrices favor cell adhesion, whereas the latter states that only the substrate with an elasticity similar to that of the microenvironment that cells experience in vivo can facilitate their maturation and function. In addition, a number of recent studies have reported that cells would modify their stiffness by cytoskeleton remodeling in response to substrates with different stiffness (Solon, et al., 2007; Abidine, et al., 2018; Jannatbabaei, 2018). For instance, cells cultured on a soft substrate exhibit a considerably lower Young's modulus than those cultured on a stiff one (Solon, et al., 2007; Abidine, et al., 2018; Jannatbabaei, 2018). Therefore, an interesting open question arises as to why cells, on one hand, favor a stiff substrate and, on the other hand, tend to lower their stiffness to adapt to a soft matrix.

A number of theoretical models have been proposed to examine the collective behavior of cell adhesion bonds since the pioneering work of Bell (1978), who first introduced mechanical forces into biological processes and put forward a critical force as a criterion to judge whether adhesion bonds are stable or not. After their seminal works, Erdmann and Schwarz (2004; 2006) further developed a more rigorous theory based on the stochastic description of molecular bonds to extract the lifetime and strength of FAs by a cluster of parallel bonds under constant load. A theoretical model involving the elasticity of the cell/substrate was established by Qian et al (2008) to qualitatively understand how cells interact with their surroundings in vivo and in vitro. This model integrated elastic effects into dissociation processes through the stress concentration resulting from asymmetrical deformation of elastic bodies. Moreover, this model predicted a crack-like failure behavior of FAs under strong stress concentration.

The stability of FAs depends on two chemomechanical processes (Qian et al., 2008; Gao, et al., 2011; Chen et al., 2015). The first process involves dissociation events that are affected by forces acting on adhesion bonds. Experiments have shown that a single adhesion bond has a binding energy of only ~10–25 $k_B T$ (Leckband and Israelachvili, 2001), which leads to a finite lifetime even in the absence of external load. When a bond is stressed by a force, the dissociation rate of the bond increases exponentially with the force due to the reduction of its thermally activated energy barrier that dominates the dissociation behaviors, as theoretically and experimentally demonstrated by dynamic force spectroscopy (Evans and Ritchie, 1997; Merkel, et al., 1999; Florin, et al., 1994). As one bond breaks, the force on other closed bonds increases that further enhances the probability of the bond to rupture (Erdmann and Schwarz, 2004). With regard to the elasticity of a cell/substrate, molecular bonds would dissociate more easily at the edge of FA because of the concentration of stress due to the decrease in cell–ECM modulus (Qian, et al., 2008).

The second process concerns association events that play an important role in maintaining stable FAs. The stability of FAs usually depends on the ratio of association rate to dissociation rate. However, quantitative theories on how a ligand connects to a receptor are deficient compared with the tremendous knowledge on how a closed bond breaks. When cells come across each other, the cellular membrane comes in contact with one another for the ligands to activate their receptors, and then ligand–receptor bonds form (Qian et al., 2008; Gao, et al., 2011; Chen et al., 2015). The association rate for a ligand and its receptor would be strongly dependent on surface separation as predicted in previous works (Erdmann and Schwarz, 2006; Qian et al., 2008). However, the association kinetics of adhesion molecules is also related to the escape from a thermally activated energy barrier, which can be greatly transformed by deformations of cell membrane/cytoskeleton to which adhesion molecules attach. These dependencies on surface separation and deformation of cell membrane/cytoskeleton, in essence, imply the coupling between deformability of cells/matrices and association process of ligand–receptor bonds. This coupling is usually exhibited as a phenomenon of rigidity-influenced FA dynamics because deformability associates with medium stiffness. For example, experiments have demonstrated that cells cultured on flexible matrices have difficulty forming stable FAs. Instead, cells form a more dynamic and diffusive structure called focal complexes, and they prefer to migrate to a stiffer region when the matrix has a rigidity gradient (Pelham and Wang, 1997; Lo et al., 2000; Discher et al., 2005).

Nevertheless, an explicit energy landscape for coupling the association process of adhesion molecules with the deformation of cells/matrices is difficult to determine due to the complexity of cell structures, including the fluctuation of membrane and remodeling of cytoskeleton. Hence, a stochastic–elasticity model was proposed by Qian et al. (2008) to evaluate the stability of molecular bonds. This model is based on stochastic process that deals with the stochasticity of chemical reaction processes of adhesion molecules and on continuum mechanics that reduces the complexity of cell/matrix structures. Although they succeeded in clarifying how the deformability of cells/matrices can provide an extra separation to adhesion molecules to influence their reactions, a quantitative understanding of how the deformation of the whole system and the distribution of interfacial bonds may influence the energetic landscape of chemical reactions of adhesion molecules is still lacking. Explicitly expressed and generally applicable analytical formulae are warranted to quantitatively characterize the association rates of ligand–receptor bonds anchored on soft deformable media.

Aside from theoretical studies, experimental observations and measurements are crucial toward quantitatively understanding, describing, and calibrating the matrix deformation-mediated reaction process of ligand–receptor bonds. The strength of a single molecular bond is measured using various experimental installations with different probe stiffness, such as an atomic force microscope (with typical probe stiffness values of 10–100 pN/nm), a biomembrane force probe

(0.1–1 pN/nm), and optical tweezers ($10^{-3}$–$10^{-2}$ pN/nm) (Evans and Ritchie, 1997; Walton et al., 2008; Zhang et al., 2008). Given that a low stiffness may enhance the sensitivity of various interactions between the probe and the surface, researchers usually recommend that probes are selected with comparable stiffness, or even considerably lower stiffness, than that of the samples to be measured (Evans, 2001). However, the measured strength appears to distinctly depend on the stiffness of loading devices following this suggestion (Walton et al., 2008; Zhang et al., 2008; Evans, 2001). Under the same loading rate, a probe with a low stiffness usually associates with a low measured strength. These stiffness-dependent strength measurements experimentally reflect the coupling between deformation of the loading devices and disassociation/association processes of molecular bonds. How to quantitatively describe the coupling of the stiffness of deformable media (cell/substrate, loading device) with the stochastically chemical reaction of bonds is crucial in understanding the biomechanical interactions of cells with their environments.

In this paper, we first derived a general description for the elasticity-associated rebinding rate of ligand–receptor bonds by rigorously considering the elastic energy of the whole system and treating the reaction of each bond as an event of Langevin dynamics. We then applied this rebinding rate to understand FA stability.

## 2. Model Description

### 2.1 General description of elasticity-associated chemical reaction of ligand–receptor bonds.

We consider the adhesion of two plane-strain soft elastic media via a cluster of ligand–receptor bonds, as shown in Fig. 1*A*. For simplicity, we may assume that the ligands are anchored to the surface of the upper medium via a polymer tethering and the receptors to that of the lower medium, which are only effectively considered as point-like binding sites. As shown in Fig. 1*A*, a longitudinal coordinate is placed at the center of the interface so that $n$ ligand–receptor pairs exist at positions $x_i$, $i = 1, 2, …, n$. We assume that the $m$th ligand–receptor pair is in an open state. A fixed vertical $z$-coordinate is placed at the binding site of the free ligand, whose equilibrium state corresponds to $z = 0$. The $z$-coordinate of the corresponding receptor usually depends on the deformation of the ligand. We denote its $z$-coordinate as $\delta(x_m)$ when the ligand is in the equilibrium state or coordinate $z = 0$.

Transition of the $m$th ligand–receptor bond from open to closed states (Qian et al., 2008) requires the ligand and the receptor to come sufficiently close within a reaction zone to form a complex and then to react at a rate $k_{on}^0$ to form a closed bond. During this process, the end of the ligand undergoes thermal fluctuation within a potential field $U(z)$, which can be viewed as a function of its vertical position $z$ where fluctuation along other directions is neglected. As adhesion molecules are linked in various ways to the interfacial structure rooted in the soft adhesion media, the potential $U(z)$ should thus include the change in deformation energies stored in the media and all other closed ligand–receptor bonds. The work of the external forces is as follows:

$$U(z) = U_e(z) - U_e(0) - W(z), \quad z \in (-\infty, \delta'(x_m)], \tag{1}$$

where $U_e(z) - U_e(0)$ represents the deformation energy change; $W(z)$ is the work done by external loads when the position of the ligand changes from 0 to $z$; and $\delta'(x_m)$ is the binding distance from the free ligand to its receptor, which represents the actual distance that the end of the free ligand should travel to meet the receptor $\delta'(x_m) = \delta(x_m) + r(x_m)$, in which $r(x_m)$ is the extra displacement of the $m$th receptor when it is touched by the $m$th ligand, as shown in Fig. 1C. $\delta'(x_m)$ becomes equal to $\delta(x_m)$ only when the adhesion media are rigid so that $r(x_m)=0$. Otherwise, $U(z) = \infty$.

We define $p(z, t)$ as the probability density to find the free ligand at position $z$ and time $t$. We have the following Smoluchowski equation to describe the statistics of the ligand (Erdmann and Schwarz, 2007; Kramers, 1940; Hanggi et al., 1990):

$$\frac{\partial p(z,t)}{\partial t} = \frac{\partial}{\partial z}\left(\frac{D}{k_B T}\frac{dU(z)}{dz} + D\frac{\partial}{\partial z}\right)p(z,t), \tag{2}$$

where $D$ is the diffusion coefficient, the value of which depends on the properties of the ligand and the solution; $k_B$ is the Boltzmann constant; and $T$ is the absolute temperature.

As the free ligand reaches thermal equilibrium within the potential field, its steady probability distribution can be easily calculated (Erdmann and Schwarz, 2007) as

$$p_{eq}(z) = \frac{1}{Z}\exp(-U(z)/k_B T), \tag{3}$$

where $Z = \int_{-\infty}^{\delta'(x_m)}\exp(-U(z)/k_B T)dz$ is the partition function. The probability of the free ligand to come sufficiently close to the receptor to form a complex within a reaction zone $[\delta'(x_m) - l_{bind}, \delta'(x_m)]$ is $\int_{\delta'(x_m)-l_{bind}}^{\delta'(x_m)} p_{eq}(z)dz \approx l_{bind} p_{eq}(\delta'(x_m))$, where $l_{bind}$ is the binding radius. By multiplying it by the spontaneous reaction rate $k_{on}^0$, we finally obtain the rebinding rate (Erdmann and Schwarz, 2006; Qian et al., 2008) as

$$k_{on} = k_{on}^0 l_{bind} \frac{\exp[-U(\delta'(x_m))/k_B T]}{\int_{-\infty}^{\delta'(x_m)} \exp(-U(z)/k_B T) dz}. \tag{4}$$

**2.2 Adhesion of rigid media.**

As an illustrative example, we consider the adhesion of two rigid bodies via $n$ closed and 1 open ligand–receptor bonds under displacement (Wang and Huang, 2015) and force-controlled loading (Li et al., 2016), as shown in the Fig. 2 *A* and *B*. By modeling each ligand as a Hookean spring with stiffness $k_{LR}$ and rest length $l_b$, we derive the rebinding rate of the open ligand–receptor bond as follows.

For the adhesion under displacement-controlled loading, a constant interfacial separation $h_0$ is maintained, as shown in Fig. 2*A*. $z = 0$ corresponds to the undeformed state of the free ligand. The elastic energy stored in all bonds at $z = 0$ can be expressed as $U_e(0) = nk_{LR}(h_0-l_b)^2/2$. When the end of the free ligand fluctuates to position $z$, the ligand extension is also $z$, and this energy becomes $U_e(z) = nk_{LR}(h_0-l_b)^2/2 + k_{LR}z^2/2$. We have the energy change as $U_e(z) - U_e(0) = k_{LR}z^2/2$. In this case, the work of the external forces is zero as the interfacial separation remains constant. Eventually, we have the potential field for the free ligand in Eq. (1) as

$$U(z) = k_{LR} z^2 / 2, \quad z \in [-l_b, h_0 - l_b], \tag{5}$$

where $\delta'(x_m) = \delta(x_m) = h_0 - l_b$, $x_m$ denotes the position of the open bond. The potential energy $U(z)$ is independent of the binding states of other bonds. Otherwise, $U(z) = \infty$. By substituting Eq. 5 into Eq. (4), we obtain

$$k_{on} = k_{on}^0 l_{bind} \sqrt{\frac{2k_{LR}}{\pi k_B T}} \frac{\exp[-k_{LR}(h_0-l_b)^2/2k_B T]}{\text{Erf}[\sqrt{k_{LR}(h_0-l_b)^2/2k_B T}] + \text{Erf}[\sqrt{k_{LR}l_b^2/2k_B T}]}, \tag{6}$$

where Erf($x$) is error function. Eq. (6) is identical to the previous result by Erdmann and Schwarz (2006).

For the adhesion under force-controlled loading shown in Fig. 2*B*, a constant pulling force $F$ is applied to the system. The elastic energy stored in the bonds at $z = 0$ is $U_e(0) = F^2/2nk_{LR}$, and the extension of the closed bonds is $h = F/nk_{LR} + l_b$. When the end of the free ligand is at position $z$, its extension can be estimated as $z_e = z - (h-h')$, where $h' = (F-k_{LR}z_e)/nk_{LR} + l_b$ is the new equilibrium interfacial separation due to the extension of the free ligand. Then, the ligand extension can be derived as $z_e = z/(1+1/n)$, which obviously depends on the number of closed bonds $n$. The elastic energy stored in all bonds becomes $U_e(z) = (F-k_{LR}z_e)^2/2nk_{LR} + k_{LR}z_e^2/2 = F^2/2nk_{LR} - Fz/(1+n) + nk_{LR}z^2/2(1+n)$. The work done by the external force due to the extension of the free ligand is $W(z) = -F(h-h') = -Fz/(1+n)$. The negative work means the system energy is increased. In this case, we can easily identify that $\delta'(x_m) = \delta(x_m) = F/nk_{LR}$. By substituting the above energy contributions to Eq. (1), we obtain the overall potential energy as a function of $z$ as

$$U(z) = nk_{LR}z^2/2(1+n), \quad z \in (-\infty, F/nk_{LR}], \tag{7}$$

which reflects the energetic landscape applied to the fluctuating free ligand. Then, the rebinding rate becomes

$$k_{on} = k_{on}^0 l_{bind} \sqrt{\frac{2k_{LR}}{\pi k_B T}} \sqrt{\frac{n}{n+1}} \frac{\exp[-F^2/2n(n+1)k_{LR}k_B T]}{1+\mathrm{Erf}[\sqrt{F^2/2n(n+1)k_{LR}k_B T}]}. \tag{8}$$

These examples clearly show that the rebinding rate of an open bond is related to the elastic energy of the whole system and the binding states of all other bonds.

### 2.3 Rupture and rebinding rates of ligand–receptor bonds between two elastic half-spaces.

The theoretical model is shown in Fig. 1*A*. We assume that the ligands are anchored to the surface of the upper medium via polymer tethering with spring constant $k_{LR}$ and rest length $l_b$ and the receptors to that of the lower medium. The total number of molecular bonds is $N_t$ evenly distributed within $2a$-width region with bond spacing $b$. The positions of $N_t$ bonds can be denoted as $\{\mathbf{x}\}$, where $\{\mathbf{x}\} = \{x_1, x_2, \ldots, x_k, \ldots, x_{N_t}\}^T$. We prescribe $N_t = 2a/b$ and $x_k = -a + (2k-1)b/2$. At the initial time, we assume that the number of closed bonds is $n$. The positions of the $n$ closed bonds are selected randomly, and we denote them as $\{\mathbf{x}_{close}\}$, where $\{\mathbf{x}_{close}\} = \{x_i \in \{\mathbf{x}\}\}_{n \times 1}$. The subset $\{\mathbf{x}_{close}\}$ not only records the closed bonds' positions, $x_i$, but also the total number of closed/open bonds, $n = $ size of $\{\mathbf{x}_{close}\}$. Accordingly, the positions of open bonds can be denoted as $\{\mathbf{x}_{open}\}$, where $\{\mathbf{x}_{open}\} = \{x_m \in \{\mathbf{x}-\mathbf{x}_{close}\}\}_{(N_t-n) \times 1}$. The elements in $\{\mathbf{x}_{close}\}$ and $\{\mathbf{x}_{open}\}$ are arranged alphabetically. To simplify, we still use symbols $x_i$, $x_j$ to denote the positions of closed bonds and symbol $x_m$ to represent the counterpart of open bonds in the following text.

We first derive the forces of closed bonds at position $\{\mathbf{x}_{close}\}$ when an arbitrary free ligand at position $x_m \in \{\mathbf{x}_{open}\}$ fluctuates a distance $z(x_m)$ toward its receptor. We denote the tethering extension of the fluctuating ligand as $z_{LR}(x_m,z)$. Herein, we assume that in the rebinding processes of free ligands to receptors, only the free ligand under investigation can experience effective fluctuation to bind its receptor and fluctuation of all other free ligands' tethering extension and the rest of the adhesion system counteracts each other so that $\langle z_{LR}(x_k \in \{\mathbf{x}_{open}\text{-}x_m\}) \rangle = 0$. As shown in Fig. 1C, the ligand's moving distance $z(x_m)$ technically is not equal to tethering extension $z_{LR}(x_m,z)$, as stated in the above force-loading case. The normal displacements of the upper elastic half-space $w^U(x_i,z)$ and the lower elastic half-space $w^L(x_i,z)$ at position $x_i \in \{\mathbf{x}_{close}\}$ induced by the closed bond force $F(x_j,z)$ at position $x_j \in \{\mathbf{x}_{close}\}$ can be calculated by the elastic Green function (Qian et al., 2008):

$$\begin{cases} w^U(x_i,z) = \sum_{j=1}^n G^U(x_i,x_j)F(x_j,z) + G^U(x_i,x_m)k_{LR}z_{LR}(x_m,z) \\ w^L(x_i,z) = \sum_{j=1}^n G^L(x_i,x_j)F(x_j,z) \end{cases}. \quad (9)$$

We rewrite these equations in matrix forms as

$$\begin{cases} \mathbf{w}^U(\{\mathbf{x}_{close}\},z) = \mathbf{G}^U(\{\mathbf{x}_{close}\},\{\mathbf{x}_{close}\}^T)\mathbf{F}(\{\mathbf{x}_{close}\},z) + \mathbf{G}^U(\{\mathbf{x}_{close}\},x_m)k_{LR}z_{LR}(x_m,z) \\ \mathbf{w}^L(\{\mathbf{x}_{close}\},z) = \mathbf{G}^L(\{\mathbf{x}_{close}\},\{\mathbf{x}_{close}\}^T)\mathbf{F}(\{\mathbf{x}_{close}\},z) \end{cases}, \quad (10)$$

where $\mathbf{w}^U(\{\mathbf{x}_{close}\},z) = \{w^U(x_i,z)\}_{n\times 1}$ and $\mathbf{w}^L(\{\mathbf{x}_{close}\},z) = \{w^L(x_i,z)\}_{n\times 1}$ are the vector forms of normal displacements of the upper and lower half-spaces, respectively, at positions $\{\mathbf{x}_{close}\}$ when the free ligand at position $x_m \in \{\mathbf{x}_{open}\}$ moves a distance $z$ toward its receptor. $\mathbf{F}(\{\mathbf{x}_{close}\},z) = \{F(x_j,z)\}_{n\times 1}$ is the force vector of closed bonds at position $x_j \in \{\mathbf{x}_{close}\}$. $\mathbf{G}^U(\{\mathbf{x}_{close}\},\{\mathbf{x}_{close}\}^T) = \{G^U(x_i,x_j)\}_{n\times n}$ and $\mathbf{G}^L(\{\mathbf{x}_{close}\},\{\mathbf{x}_{close}\}^T) = \{G^L(x_i,x_j)\}_{n\times n}$ are the corresponding compliance matrices for closed bonds, where $G^U(x_i,x_j) = 2(1-v_U^2)\ln|(x_\infty-x_j)/(x_i-x_j)|/\pi b E_U$ and $G^L(x_i,x_j) = 2(1-v_L^2)\ln|(x_\infty-x_j)/(x_i-x_j)|/\pi b E_L$ for $i \neq j$. For $i = j$, $G^U(x_i,x_i) = 2(1-v_U^2)C_0/\pi b E_U \triangleq G_0^U$ and $G^L(x_i,x_i) = 2(1-v_L^2)C_0/\pi b E_L \triangleq G_0^L$ (16). $C_0 = 1 + \ln(x_\infty/2a_0)$ is a constant selected to satisfy the condition that $\mathbf{F}(\{\mathbf{x}_{close}\},z)$ causes zero elastic displacements at infinity $x_\infty$, and $a_0$ denotes the radius of individual bond with the typical value on the order of a few nanometers (16). $\mathbf{G}^U(\{\mathbf{x}_{close}\},x_m) = \{G^U(x_i,x_m)\}_{n\times 1}$ is the compliance vector for an open bond at position $x_m \in \{\mathbf{x}_{open}\}$ to closed bonds at positions $x_i \in \{\mathbf{x}_{close}\}$. Note that $|x_\infty/x_i| \gg 1$, then we have

$$G^U(x_i,x_j) - G^U(x_j,x_i) = 2(1-v_U^2)\ln|(x_\infty-x_j)/(x_\infty-x_i)|/\pi b E_U \to 0, \quad (11)$$

that is, $G^U(x_i,x_j) = G^U(x_j,x_i)$. This result is suitable for $G^L(x_i,x_j) = G^L(x_j,x_i)$. Hence,

$\mathbf{G}^U(\{\mathbf{x}_{close}\},\{\mathbf{x}_{close}\}^T)$ and $\mathbf{G}^L(\{\mathbf{x}_{close}\},\{\mathbf{x}_{close}\}^T)$ are symmetrical matrices.

The elastic extension of the $i$th closed bond $u(x_i,z) = F(x_i,z)/k_{LR}$ and the related normal surface displacements $w^U(x_i,z)$ and $w^L(x_i,z)$ of the two elastic half-spaces should satisfy the equation $u(x_i,z) + w^U(x_i,z) + w^L(x_i,z) = h(z)$, where $h(z)$ is the interface separation at infinity $x_\infty$. It gives

$$F(x_i,z)/k_{LR} + \left(\sum_{j=1}^{n} G^U(x_i,x_j)F(x_j,z) + G^U(x_i,x_m)k_{LR}z_{LR}(x_m,z)\right) + \sum_{j=1}^{n} G^L(x_i,x_j)F(x_j,z) = h(z) .$$

We rewrite these equations in a matrix form as

$$\mathbf{A}(\{\mathbf{x}_{close}\},\{\mathbf{x}_{close}\}^T)\mathbf{F}(\{\mathbf{x}_{close}\},z) + \mathbf{G}^U(\{\mathbf{x}_{close}\},x_m)k_{LR}z_{LR}(x_m,z) = \mathbf{J}h(z) , \quad (12)$$

where $\mathbf{A}(\{\mathbf{x}_{close}\},\{\mathbf{x}_{close}\}^T) = \mathbf{G}^U(\{\mathbf{x}_{close}\},\{\mathbf{x}_{close}\}^T) + \mathbf{G}^L(\{\mathbf{x}_{close}\},\{\mathbf{x}_{close}\}^T) + \mathbf{I}/k_{LR}$ is a symmetrical matrix dependent on the position and number of closed bonds $\{\mathbf{x}_{close}\}$. $\mathbf{I}$ is an $n$-by-$n$ unitary matrix, and $\mathbf{J} = \{1\}_{n\times 1}$ is an $n$-by-1 vector with all elements equal to 1. The $(n+1)$ unknowns $\{\mathbf{F}(\{\mathbf{x}_{close}\},z); h(z)\}$ can be solved from Eq. (12) together with the global force balance as follows:

$$\mathbf{J}^T\mathbf{F}(\{\mathbf{x}_{close}\},z) = F - k_{LR}z_{LR}(x_m,z) , \quad (13)$$

By combining Eq. (12) with Eq. (13), we have

$$\begin{bmatrix} \mathbf{A}(\{\mathbf{x}_{close}\},\{\mathbf{x}_{close}\}^T) & -\mathbf{J} \\ \mathbf{J}^T & 0 \end{bmatrix} \begin{bmatrix} \mathbf{F}(\{\mathbf{x}_{close}\},z) \\ h(z) \end{bmatrix} = \begin{bmatrix} -\mathbf{G}^U(\{\mathbf{x}_{close}\},x_m)k_{LR}z_{LR}(x_m,z) \\ F - k_{LR}z_{LR}(x_m,z) \end{bmatrix} . \quad (14)$$

For convenience, in the following text, we denote $\mathbf{A}(\{\mathbf{x}_{close}\},\{\mathbf{x}_{close}\}^T)$ as $\mathbf{A}$ and $\mathbf{F}(\{\mathbf{x}_{close}\},z)$ as $\mathbf{F}(z)$. By solving Eq. (14) directly, we can easily obtain the forces of closed bonds $\mathbf{F}(z)$ and the interface separation $h(z)$ as follows

$$\begin{cases} \mathbf{F}(z) = \dfrac{1}{\mathbf{J}^T\mathbf{A}^{-1}\mathbf{J}}\left[\mathbf{A}^{-1}\mathbf{J}F - \left((\mathbf{J}^T\mathbf{A}^{-1}\mathbf{J}\mathbf{A}^{-1} - \mathbf{A}^{-1}\mathbf{J}\mathbf{J}^T\mathbf{A}^{-1})\mathbf{G}^U(\{\mathbf{x}_{close}\},x_m) + \mathbf{A}^{-1}\mathbf{J}\right)k_{LR}z_{LR}(x_m,z)\right] \\ h(z) = \dfrac{1}{\mathbf{J}^T\mathbf{A}^{-1}\mathbf{J}}\left[F + \left(\mathbf{J}^T\mathbf{A}^{-1}\mathbf{G}^U(\{\mathbf{x}_{close}\},x_m) - 1\right)k_{LR}z_{LR}(x_m,z)\right] \end{cases} . \quad (15)$$

At the initial state, the free ligand remains, i.e., $z_{LR}(x_m,0) = 0$, we have

$$\begin{cases} \mathbf{F}(0) = \mathbf{A}^{-1}\mathbf{J}F / \mathbf{J}^T\mathbf{A}^{-1}\mathbf{J} \\ h(0) = F / \mathbf{J}^T\mathbf{A}^{-1}\mathbf{J} \end{cases} . \quad (16)$$

After obtaining the forces assigned to closed bonds, we can easily calculate the rupture rates of closed bonds via Bell's rate equation (Bell, 1978):

$$\mathbf{k}_{off}(\{\mathbf{x}_{close}\}) = k_0 \exp[\mathbf{F}(0)/F_b] , \quad (17)$$

where $\mathbf{k}_{off}(\{\mathbf{x}_{close}\}) = \{k_{off}(x_i)\}_{n\times 1}$, $k_0$ is the spontaneous rupture rate in the absence of force, and $F_b$ is the force scale.

Now, we calculate the rebinding rates for open bonds at position $\{\mathbf{x}_{open}\}$. As shown in Fig. 1C, the relationship of tethering extension $z_{LR}(x_m,z)$ to the free ligand's motion $z(x_m)$ can be written as

$$z(x_m) = z_{LR}(x_m, z) + [w^U(x_m, z) - w^U(x_m, 0)] + [h(0) - h(z)]. \tag{18}$$

By substituting the expressions of $w^U(x_m,z)$ in Eq. (9) and $h(z)$ in Eq. (15) into Eq. (18), we obtain

$$z(x_m) = M_0(x_m) z_{LR}(x_m, z), \tag{19}$$

where

$$M_0(x_m) = 1 + k_{LR} G_0^U + k_{LR} \left( \frac{\left(\mathbf{J}^T \mathbf{A}^{-1} \mathbf{G}^U(\{\mathbf{x}_{close}\}, x_m) - 1\right)^2}{\mathbf{J}^T \mathbf{A}^{-1} \mathbf{J}} - \mathbf{G}^U(\{\mathbf{x}_{close}\}^T, x_m) \mathbf{A}^{-1} \mathbf{G}^U(\{\mathbf{x}_{close}\}, x_m) \right), \tag{20}$$

where $\mathbf{G}^U(\{\mathbf{x}_{close}\}^T, x_m) = [\mathbf{G}^U(\{\mathbf{x}_{close}\}, x_m)]^T$. Eqs. (19) and (20) analytically give the relationship of tethering extension $z_{LR}(x_m,z)$ to the motion of free ligands $z(x_m)$ at position $x_m \in \{\mathbf{x}_{open}\}$. The first term in the right hand side of $M_0(x_m)$ demonstrates the contribution of the tethering fluctuation to the ligand's motion. The second term represents the local fluctuation of the upper elastic half-space where the free ligand is located. The final term embodies the complete effects of system deformation and the binding states of other bonds.

When a free ligand at position $x_m$ moves a distance $z(x_m)$ toward its receptor, the potential energies $U_e(z)$ stored in the ligand–receptor bonds and in the two elastic half-spaces can be expressed as

$$U_e(z) = U_{LR}(z) + U_{body}(z), \tag{21}$$

where $U_{LR}(z) = \mathbf{F}^T(z)\mathbf{F}(z)/2k_{LR} + k_{LR} z_{LR}^2(x_m,z)/2$ is the elastic energies of ligand–receptor bonds stored in closed bonds at $\{\mathbf{x}_{close}\}$ and in the tethering at $x_m \in \{\mathbf{x}_{open}\}$. $U_{body}(z) = \mathbf{F}^T(z)[\mathbf{w}^U(\{\mathbf{x}_{close}\},z)+ \mathbf{w}^L(\{\mathbf{x}_{close}\},z)]/2 + k_{LR} z_{LR}(x_m,z) w^U(x_m,z)/2$ is the elastic potential energy, which is obtained by the reciprocal work theorem (31), stored in the two elastic half-spaces. The work done by the external force is $W(z) = -F[h(0)-h(z)]$. By substituting the above energy contributions to Eq. (1), we can obtain the overall potential energy as a function of $z$ as

$$U(z) = \frac{1}{2} \frac{k_{LR}}{M_0(x_m)} z^2(x_m), \quad z \in (-\infty, \delta'(x_m)]. \tag{22}$$

To calculate the actual binding distance $\delta'(x_m)$, we need to obtain the separation $\delta(x_m,z)$ of the free ligand to its receptor at position $x_m$. As shown in Fig. 1C, $\delta(x_m,z) = h(z) - w^U(x_m,z) - l_b - z_{LR}(x_m,z) - w^L(x_m,z)$. By plugging the related equations into $\delta(x_m,z)$, we obtain

$$\delta(x_m, z) = \delta(x_m, 0) - z[1 - M_1(x_m)/M_0(x_m)], \tag{23}$$

where

$$\delta(x_m,0) = \left[1 - \mathbf{J}^T\mathbf{A}^{-1}\left(\mathbf{G}^U(\{\mathbf{x}_{\text{close}}\}, x_m) + \mathbf{G}^L(\{\mathbf{x}_{\text{close}}\}, x_m)\right)\right] F / \mathbf{J}^T\mathbf{A}^{-1}\mathbf{J}. \quad (24)$$

and

$$M_1(x_m) = k_{\text{LR}} \mathbf{G}^L(\{\mathbf{x}_{\text{close}}\}^T, x_m) \left( \frac{\mathbf{A}^{-1}\mathbf{J}\left(1 - \mathbf{J}^T\mathbf{A}^{-1}\mathbf{G}^U(\{\mathbf{x}_{\text{close}}\}, x_m)\right)}{\mathbf{J}^T\mathbf{A}^{-1}\mathbf{J}} + \mathbf{A}^{-1}\mathbf{G}^U(\{\mathbf{x}_{\text{close}}\}, x_m) \right), \quad (25)$$

where $\mathbf{G}^L(\{\mathbf{x}_{\text{close}}\}^T, x_m) = [\mathbf{G}^L(\{\mathbf{x}_{\text{close}}\}, x_m)]^T$. Once the free ligand at $x_m$ fluctuates and contacts the receptor, i.e., $z(x_m) = \delta'(x_m)$, we have $\delta(x_m, \delta'(x_m)) = \delta(x_m, 0) - \delta'(x_m)[1 - M_1(x_m)/M_0(x_m)] = 0$. Upon solving this equation, we obtain

$$\delta'(x_m) = \delta(x_m, 0)\left(1 + \frac{M_1(x_m)}{M_0(x_m) - M_1(x_m)}\right) = \delta(x_m, 0) + r(x_m), \quad (26)$$

where $r(x_m) = \delta(x_m,0)M_1(x_m)/[M_0(x_m)-M_1(x_m)]$. Now, we have obtained the truncated potential field $U(z)$ and the binding distance $\delta'(x_m)$. By plugging these quantities into Eq. (4), we finally obtain the rebinding rate for the free ligand to its receptor at position $x_m \in \{\mathbf{x}_{\text{open}}\}$:

$$k_{\text{on}}(x_m) = k_{\text{on}}^0 l_{\text{bind}} \sqrt{\frac{2k_{\text{LR}}}{\pi M_0(x_m)k_B T}} \frac{\exp[-k_{\text{LR}}\delta'(x_m)^2/2M_0(x_m)k_B T]}{1 + \text{Erf}[\sqrt{k_{\text{LR}}\delta'(x_m)^2/2M_0(x_m)k_B T}]}. \quad (27)$$

Several examples are provided for further analysis, and parameter values unspecified in the text, such as $k_{\text{LR}}$, and $l_b$, are selected according to Table 1.

## 3. Results

### 3.1 Rebinding rate in a two-bond system wherein one bond breaks.

A two-bond system wherein one bond breaks at position $x_1 = -b/2$ and the other bond is closed at position $x_2 = b/2$ is evaluated. In this system, the rebinding rate of the open bond can be expressed as

$$k_{\text{on}}(x_1) = k_{\text{on}}^0 l_{\text{bind}} \sqrt{\frac{2k_{\text{LR}}}{\pi M_0(x_1)k_B T}} \frac{\exp\left[-B_0 k_{\text{LR}}\delta^2(x_1,0)/2k_B T\right]}{1 + \text{Erf}[\delta(x_1,0)\sqrt{B_0 k_{\text{LR}}/2k_B T}]}, \quad (28)$$

where $M_0(x_1) = 2 + 2k_{\text{LR}}[G_0^U - G^U(x_1,x_2)] + k_{\text{LR}}G_0^L$, $\delta(x_1,0) = F\left(1/k_{\text{LR}} + G_0^U - G^U(x_1,x_2) + G_0^L - G^L(x_1,x_2)\right)$, and $B_0 = M_0(x_1)/[M_0(x_1) - k_{\text{LR}}G^L(x_1,x_2)]^2$. For comparison, we quote herein the previous rebinding rate proposed by Qian et al. (2008):

$$k_{\text{on}}^{\text{Qian}}(x_m) = k_{\text{on}}^0 l_{\text{bind}} \sqrt{\frac{2k_{\text{LR}}}{\pi k_{\text{B}} T}} \frac{\exp\left[-k_{\text{LR}}\delta^2(x_m,0)/2k_{\text{B}}T\right]}{\text{Erf}[l_b\sqrt{k_{\text{LR}}/2k_{\text{B}}T}]+\text{Erf}[\delta(x_m,0)\sqrt{k_{\text{LR}}/2k_{\text{B}}T}]}, \quad (29)$$

where $x_m = x_1$ in this two-bond system. The main difference between the new and the previous rebinding rates is that the new rebinding rate considers the deformation of the overall system and is, therefore, more sensitive to thermal fluctuations. By contrast, the previous rebinding rate considers the separation between ligands and receptors.

Qualitatively, the new rebinding rate is considerably dependent on the stiffness of the lower half-space in the absence of force $F$, as shown in Fig. 3$A$. This phenomenon is consistent with our hypothesis. However, the previous rebinding rate remains constant regardless if the lower half-space is soft or not because it is separation dependent (Qian et al., 2008). As the force approaches zero, the ligand comes close to its receptor and the separation between them remains constant. Hence, a constant rebinding rate is obtained. Quantitatively, as the external loading $F$ slightly increases, the separation of free ligands from their receptors substantially increases in the context of soft substrates. Therefore, the previous rebinding rate dramatically decreases, as shown in Fig. 3$B$. This result is not suitable for the new rebinding rate because it is energy dependent. The minute changes in external loadings cannot lead to any apparent variation in system energies.

Another characteristic of the new rebinding rate is its asymmetrical dependence on the stiffness of upper and lower half-spaces, as seen directly from Eq. (28). As shown in Fig. 4, the rebinding rate of the open bond always increases with the modulus of the lower half-space. However, once the stiffness of the substrates is given, only the upper half-space with a stiffness similar to that of the lower half-space can reach the maximum rebinding rate. This asymmetrical dependence rests on the assumption that the receptors are immobilized on the surface of the lower half-space and the rebinding processes are dependent on the ligands' fluctuations.

In this two-bond system, as the modulus of the lower half-space approaches zero, i.e., $E_{\text{L}} \to 0$ & $E_{\text{U}}/E_{\text{L}} \gg 1$, we obtain the asymptotic expression for the open bond's rebinding rate as follows:

$$k_{\text{on}}(x_1) \to k_{\text{on}}^0 l_{\text{bind}} \sqrt{2L_0 E_{\text{L}}/\pi k_{\text{B}}T} \exp[-F^2/2L_0 E_{\text{L}} k_{\text{B}}T] / \left(1 + \text{Erf}[\sqrt{F^2/2L_0 E_{\text{L}} k_{\text{B}}T}]\right), \quad (30)$$

where $L_0 = \pi b/2C_0(1-v_{\text{L}}^2)$. Note that $\text{Erf}[\sqrt{F^2/2L_0 E_{\text{L}} k_{\text{B}}T}] \to 1$ in the context of non-zero force $F$ (i.e. $F^2 \gg E_{\text{L}}$). Eq. (30) states that the rebinding rate of the open bond is sensitive to the modulus $E_{\text{L}}$ and the external force $F$, that is, $k_{\text{on}}(x_1) \propto E_{\text{L}}^{1/2}\exp[-c(F^2)/E_{\text{L}}]$, where $c(F^2) = F^2/2L_0 k_{\text{B}}T$. As the force

approaches zero, the rebinding rate only depends on the modulus of the lower elastic half-space:

$$k_{on}(x_1) \to k_{on}^0 l_{bind} \sqrt{2L_0 E_L / \pi k_B T} \ . \tag{31}$$

We emphasize that the asymptotic expressions in Eqs. (30) and (31) are not only suitable for the two-bond system but also applicable for molecular clusters of multiple bonds. We just need to substitute $L_0$ with $\mathbf{J}^T \mathbf{L}^{-1} \mathbf{J}$, where $\mathbf{L} = \mathbf{G}^L(\{\mathbf{x}_{close}\},\{\mathbf{x}_{close}\}^T) \times E_L$. The softer the lower elastic body, the more fluctuation the system experiences, and the lower the rebinding rate will be. We deduce for the first time the scaling law for the rebinding rate of ligand–receptor bonds in the context of soft substrates. These asymptotic results are shown in the Fig. 5.

### 3.2 Rebinding rates of molecular bonds between rigid bodies.

As $E_U$, $E_L \to \infty$, we reduce to the situation of molecular bonds between rigid bodies. The rebinding rates for open bonds at position $\{\mathbf{x}_{open}\}$ are identical and can be expressed as

$$k_{on} = k_{on}^0 l_{bind} \sqrt{\frac{2k_{LR}}{\pi k_B T}} \sqrt{\frac{n}{n+1}} \frac{\exp[-F^2/2n(n+1)k_{LR}k_B T]}{1+\mathrm{Erf}[\sqrt{F^2/2n(n+1)k_{LR}k_B T}]} \ . \tag{32}$$

In this case, only the bonds' number $n$ plays a role, and the information of the bonds' position is lost. This result is the same as that of the force-loading situation. The main differences between the rebinding rate in Eq. (32) with that of the previous one are discussed above. Here, we complement an asymptotic expression of Eq. (32) with the external loading $F$ approaching zero, which is

$$k_{on} \to k_{on}^0 l_{bind} \sqrt{2k_{LR}/\pi k_B T} \sqrt{n/(n+1)} \ . \tag{33}$$

Eq. (33) reflects the influences of the bonds' binding state on the rebinding process. The presence of few closed bonds leads to more fluctuations in the system and lowers the rebinding rate.

### 3.3 Stochastic description and Monte Carlo simulation for the bonds' rupture and rebinding processes.

The stochastic process of the adhesion cluster of ligand–receptor bonds under an external loading can be assumed as a Markov process, as described by the following one-step master equation (Erdmann and Schwarz, 2004; Erdmann, Schwarz, 2006; Qian et al., 2008):

$$\frac{dp_n(\tau)}{d\tau} = r_{n+1}p_{n+1}(\tau) + g_{n-1}p_{n-1}(\tau) - [r_n + g_n]p_n(\tau) , \tag{34}$$

where $p_n(\tau)$ is the probability of $n$ bonds to be closed at a given dimensionless time $\tau = k_0 t$.

$r_n = \sum_{i=1}^{n} k_{\text{off}}(x_i)/k_0$ is the normalized rupture rate of transition from $n$ to $n-1$ closed bonds. $g_n = \sum_{m=1}^{N_t-n} k_{\text{on}}(x_m)/k_0$ is the normalized rebinding rate of transition from $n$ to $n+1$ closed bonds. Eq. 34 is only suitable for the spatial and temporal evolution of molecular bonds with identical rupture and rebinding rates at any given binding state $n$.

For elastic half-spaces, such as in the present case, the master equation in Eq. (34) is not applicable. Monte Carlo simulation based on the first reaction method, which is derived from the Gillespie algorithm, can be performed to solve the evolution of molecular bonds (Erdmann and Schwarz, 2004; Erdmann, Schwarz, 2006; Qian et al., 2008). The basic idea of such simulations is to cast stochastic trajectories for cluster evolution in accordance with the aforementioned reaction rates and average over many independent trials to obtain useful statistical information (Qian et al., 2008). The solving procedure is outlined as follows:

   1). The adhesion sizes $2a$ are inputted. The number of bonds is $N_t = 2a/b$. The position of bonds is $x_k = -a + (2k-1)b/2$, for $k = 1, 2, 3, \ldots, N_t$. In the beginning $\tau = 0$. The number of closed bonds is set as $n = N_t$.

   2). The positions of closed and open bonds are located and denoted as $\{\mathbf{x}_{\text{close}}\}$ and $\{\mathbf{x}_{\text{open}}\}$, respectively.

   3). The forces $\mathbf{F}(\{\mathbf{x}_{\text{close}}\},0)$ assigned to closed bonds are solved using Eq. (16). The forces are plugged into Eq. 17, and the rupture rates for the each closed bond $\mathbf{k}_{\text{off}}(\{\mathbf{x}_{\text{open}}\})$ are obtained.

   4). The normalized parameter $M_0(x_m)$ and binding distance $\delta'(x_m)$ are solved using Eqs. 20 and 26, respectively. $M_0(x_m)$ and $\delta'(x_m)$ are substituted into Eq. 27, and the rebinding rates for open bonds $k_{\text{on}}(\{\mathbf{x}_{\text{open}}\})$ are obtained.

   5). The reaction times at individual bond locations $x_k$ are calculated by (14–16) $d\tau(x_k) = -\ln[\xi(x_k)]/\mu(x_k)$, where $\xi(x_k)$ is the generated series of independent random numbers for individual reaction sites, which are uniformly distributed over the interval [0,1] and $\mu(x_k)$ is the normalized reaction rate depending on the bond states at location $x_k$ as $\mu(x_i \in \{\mathbf{x}_{\text{close}}\}) = k_{\text{off}}(x_i)/k_0$ and $\mu(x_m \in \{\mathbf{x}_{\text{open}}\}) = k_{\text{on}}(x_m)/k_0$.

   6). The smallest $d\tau(x_\mu)$ and the corresponding reaction site $x_k = x_\mu$ are recorded as the time and location associated with the next bond reaction, respectively.

   7). The bond state at site $x_\mu$ is changed. The bond status at site $x_\mu$ is changed to open if it is currently closed and to close if it is open. $\tau = \tau + d\tau(x_\mu)$ is set.

**8).** Step **2)** is repeated and looped until all bonds are open. The final lifetime is recorded as the lifetime of the cluster of molecular bonds in one trajectory.

### 3.4 Cells with a stiffness similar to that of ECM can reach the maximum adhesion strength.

In Fig. 6, we depict the normalized lifetime of a cluster of ligand–receptor bonds under different apparent stress $F/2ab$ by Monte Carlo simulation. As shown in Fig. 6, the cluster lifetime substantially decreases with apparent stress. The large stress corresponds to a quick decline in the cluster lifetime. As $F/2ab$ is below a critical value $F_{cr}/2ab$, the cluster reach a stable state, implying that its lifetime asymptotically approaches infinity. We define this $F_{cr}/2ab$ as the cluster strength (Qian et al., 2008). In numerical simulations, we determine that a cluster is stable if its survival time $\tau_\infty$ exceeds 100 (Qian et al., 2008) and define the corresponding load as the cluster strength. As depicted in Fig. 6, even if we choose $\tau_\infty = 200$ as the stable state for the cluster, the change in the external force is less than 2%. Therefore, we assert that a different selection of $\tau_\infty$ will lead to only a minor change in the predicted strength values.

Fig. 7 shows the cluster strength with respect to the modulus of the lower or upper half-spaces. As shown in Fig. 7*A*, the cluster strength always increases with the stiffness of the lower half-space regardless if the upper half-space is soft or not. This result may confirm the assumption that stiff substrates favor cell adhesion, consistent with that obtained by previous experiments (Pelham and Wang, 1997; Lo et al., 2000; Discher et al., 2005). Fig. 7*B* indicates that the molecular cluster can reach the maximum strength only when the stiffness of the upper half-space is comparable with that of the lower half-space given a specified stiffness of the lower half-space. These two results may explain the phenomena observed in previous experiments. For example, **1)** the propensity of cells to migrate from a soft area to a stiff one when cultured on substrates with different rigidity gradients is due to cells having a higher strength on stiffer substrates, as shown in Fig. 7*A* (3). **2)** The tendency of cells to lower their modulus by cytoskeleton remodeling in response to soft substrates is explained by the observation that only cells with a stiffness similar to that of substrates can achieve the maximum adhesion strength, as shown in Fig. 7*B* (Solon et al., 2007; Abidine et al., 2018; Jannatbabaei et al., 2018).

### 3.5 Cells gradually lose adhesion ability as ECM stiffness decreases.

Another interesting phenomenon depicted in Fig. 7*A* is that a very soft substrate (near 0.01 kPa) cannot afford cells to maintain a stable state (normalized lifetime lower than 100) regardless of how small the external force is. The explanation for this phenomenon, in the context of a soft substrate (say $E_L \to 0$), is that the rebinding rates of adhesion molecules tend to $k_{on} \propto (E_L)^{1/2} \exp[-c(F^2)/E_L]$ (Eq. (30)), which is considerably small to allow a breaking bond to rebind once it ruptures. This property cannot be obtained from previous models. To gain a better understanding of the unstable state of FAs on soft substrates, we plot the cluster-normalized time as a function of the external force in Fig. 8 by using the newly proposed rebinding rate (Eq. (27)) and the previous one proposed by Qian et al. (Eq. (29)).

In Fig. 8, the main difference between the new and the previous rebinding processes is exposed. In the previous rebinding processes, the bonds' rebinding rates only depend on **1)** the separation of ligands to receptors and **2)** their fluctuation. As long as the external force is sufficiently small, the cluster always has a stable state regardless of how soft the substrate is (Fig. 8*A*, black circle line). By contrast, we consider in the new rebinding process the coupling effects of deformation/fluctuation of the overall system on the energy landscape of adhesion molecules. Even in the absence of an external force, the thermal fluctuations of the overall system are remarkable on soft substrates, resulting in unstable state of the molecular cluster. This finding is consistent with the experimental observation that cells plated on soft substrates do not spread because of a sixfold thermal fluctuation around their periphery compared with those cultured on a stiff substrate (Pelham and Wang, 1997). A recent experiment (Oakes, 2018) has demonstrated that cells cannot spread on soft substrates because of the poor ability of integrins, which are prototypes of ligand–receptor bonds scattered on cell surface, to bind to their counterparts. After the addition of $Mn^{2+}$, which can specifically change the affinity of integrins, cells spread not only on soft substrates but also on stiff ones. This experiment strongly confirms our findings in Fig. 8 that the inability of cells to spread on soft substrates is also a result of their poor rebinding probability and not only of stress concentration.

### 4. Discussion

By adhering to ECM via a cluster of transmembrane proteins from the integrin family, cells can perceive mechanical cues from their surroundings and turn them into biological signals they can understand. This process is called mechanotransduction, which is crucial for many cellular biochemical functions, such as spreading, proliferation, and migration (Martino et al., 2018). Recent experiments have demonstrated that the spread of cell on a soft matrix is prohibited by strong thermal fluctuation (Pelham and Wang, 1997; Lo et al., 2000; Discher et al., 2005). However, after the addition of $Mn^{2+}$ (which specifically enhances the affinity of integrin), cells spread to an area of a soft substrate comparable with that on a stiff matrix (Oakes, 2018). These findings imply that integrin-dependent rebinding processes may be critical in cell adhesion or at least is not less important than rupture processes. Unfortunately, the importance of bonds' rebinding processes has not received as much attention as that of rupture processes.

Maintaining a maximum adhesion strength appears as the strategy of normal cells in interacting with ECMs. However, this strategy is inappropriate for cancer cells. Stiffening in ECMs is usually associated with the onset of degenerative diseases, such as tumor. Stiff substrates are considered to favor cancer cell invasion through the basement membrane of local tissues to migrate to other organs (Martino et al., 2018). Accumulating evidence indicates that cancer cells have a smaller stiffness than normal cells when cultured on a Petri dish. Recent studies have reported that cells exhibit a smaller modulus on stiff substrates and a larger modulus on soft substrates compared with their normal counterparts (Rianna and Radmacher, 2017; Rianna, et al., 2018). These findings are against the general observation that cancer cells are softer than normal cells. The larger/smaller stiffness on soft/stiff substrates exhibited by cancer cells compared with normal cells remains poorly understood. Clearly, the rule for normal cells does not fit for cancer cells; hence, a different strategy must be considered by cancer cells.

In this regard, we replot the adhesion strength shown in Fig. 7 as a function of combined modulus $E^*$ ($1/E^* = (1-v_U^2)/E_U + (1-v_L^2)/E_L$) displayed in Fig. 9. As depicted in Fig. 9, a high substrate stiffness corresponds to a strong adhesion strength at a certain combined modulus, a result that once again confirms the assertion that stiff substrates favor cell adhesion. On one hand, maintaining a constant adhesion strength might be a feasible strategy for cancer cells to prevent them from sticking tightly to ECMs, a process that halts their invasion. On the other hand, this action protects their proliferation against being washed away by blood flow. Fig. 9 shows that maintaining a constant strength enables cancer cells to adopt various methods, such as a soft substrate with a stiff cell (arrow 1), a stiff substrate with a soft cell (arrow 2), and a moderate stiffness of cell and substrate (arrow 3). These results may provide the answer to the question as to why cancer cells become softer than normal cells on stiff substrates but stiffer on soft substrates.

## 5. Conclusion

In this study, we developed a generally analytical expression for the rebinding rates of adhesion molecules by calculating the coupling of deformability of cells/matrices with the rebinding process of bonds. The chemical reaction rates of adhesion molecules usually involve the escape from a thermally activated energy barrier, which can be greatly transformed by the deformation of cells/matrices as discussed above. This coupling involves medium stiffness and is usually related to various rigidity-dependent properties of cells. For example, a number of experimental findings can be satisfactorily explained by the present model using the new rebinding rate. We emphasize that the rebinding processes of molecular bonds is not dependent on separation but related to overall energy. Few closed bonds or soft substrates are associated with a high thermal fluctuation, resulting in low rebinding rate. This newly proposed rebinding rate may substantially improve our understanding of how a ligand attaching to the cell surface interacts with its receptor immobilized on the substrate.

In evaluating FA stability, we discovered two distinct strategies employed by normal and cancer cells. Maintaining a maximum strength favors the physiological processes of normal anchorage-dependent cells, whereas sustaining a constant strength may be beneficial for cancer cells to balance invasion and proliferation. These findings may provide a good understanding of the mechanical responses of normal and diseased cells to different stiffness matrices and may be applied to cancer diagnosis.

**Acknowledgement**

This research is supported by grants from the National Natural Science Foundation of China (11925204).

**Figures and Tables**

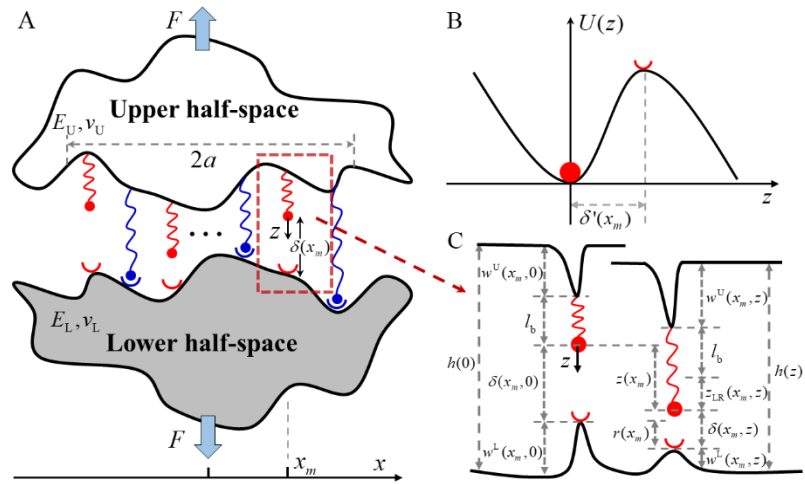

**Fig. 1.** Schematic description of two soft media interacting with each other. (*A*) The interaction between two elastic half-spaces via a cluster of ligand–receptor bonds under a force *F*. The upper elastic half-space represents the cell, whereas the lower elastic half-space represents the ECM. The bond cluster size is 2*a*. Young's modulus and Poisson's ratio for the upper and lower half-spaces are $E_U$, $\nu_U$ and $E_L$, $\nu_L$, respectively. (*B*) Potential field for the free ligand at position $x_m \in \{\mathbf{x}_{open}\}$. (*C*) Changes in interfacial displacement as the free ligand located at $x_m \in \{\mathbf{x}_{open}\}$ comes close to its receptor.

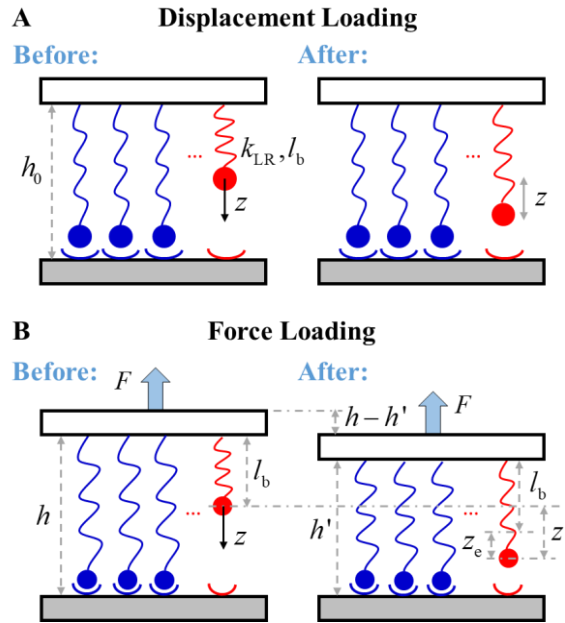

**Fig. 2.** Schematic description of a cluster of molecular bonds between two rigid bodies. The ligands attach to the upper rigid body through a linear tethering with the compliance $k_{LR}$ and rest length $l_b$, and the receptors are immobilized on the lower rigid body. (*A*) Displacement loading: the initial separation between the rigid bodies is set as $h_0$. $U(z) = k_{LR}z^2/2$. (*B*) Force loading: a force $F$ is applied to the system. $U(z) = nk_{LR}z^2/2(1+n)$.

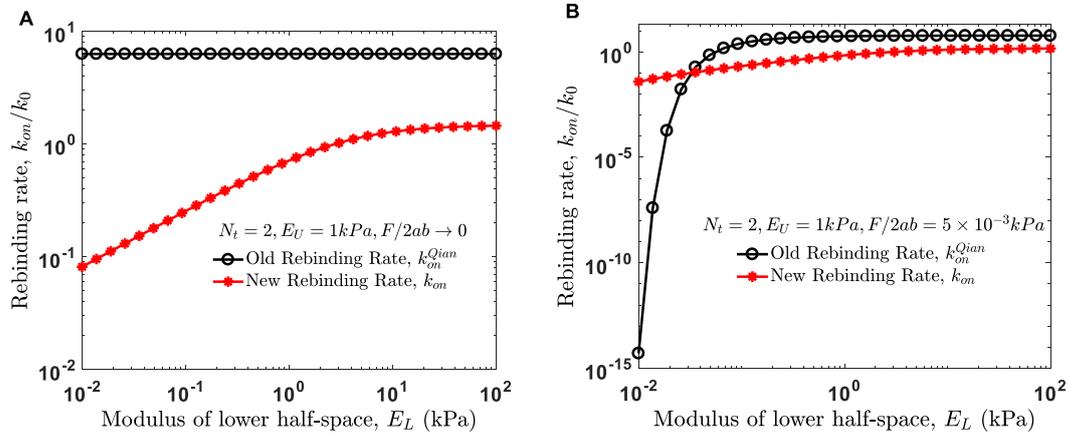

**Fig. 3.** Rebinding rate of the open bond at position $x_1$ as a function of Young's modulus of the lower half-space under different external forces. (*A*) $F/2ab \to 0$. (B) $F/2ab = 5\times10^{-3}$ kPa.

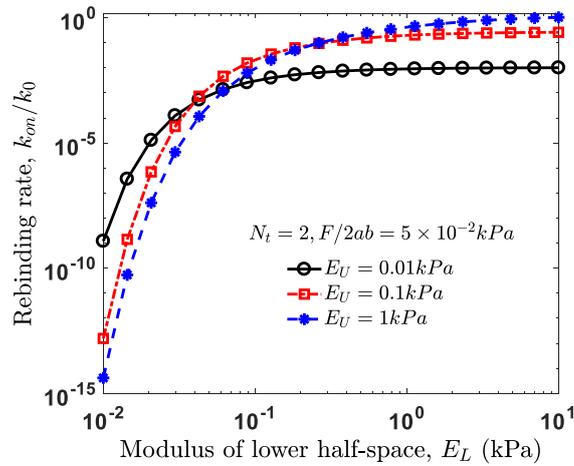

**Fig. 4.** Rebinding rate of the open bond at position $x_1$ as a function of Young's modulus of the lower half-space $E_L$ with different rigidity upper half-space.

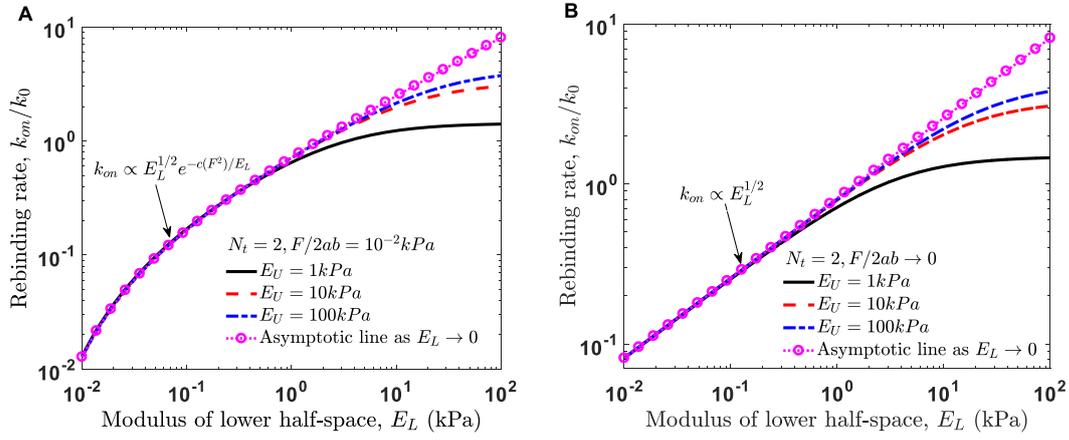

**Fig. 5.** Asymptotic expressions of the rebinding rate as a function of Young's modulus of the lower half-space $E_L$. (*A*) $F/2ab = 10^{-2}$ kPa. (*B*) $F/2ab \to 0$.

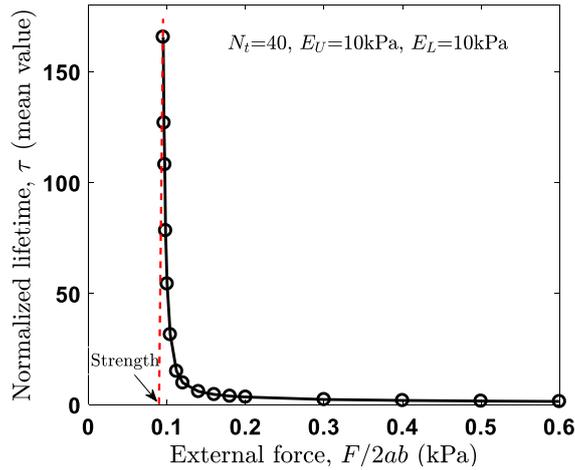

**Fig. 6.** Normalized lifetime of a cluster of ligand–receptor bonds as a function of external forces. The initial bond number is $N_t = 40$, and the modulus of the upper and lower half-spaces is set at 10 kPa.

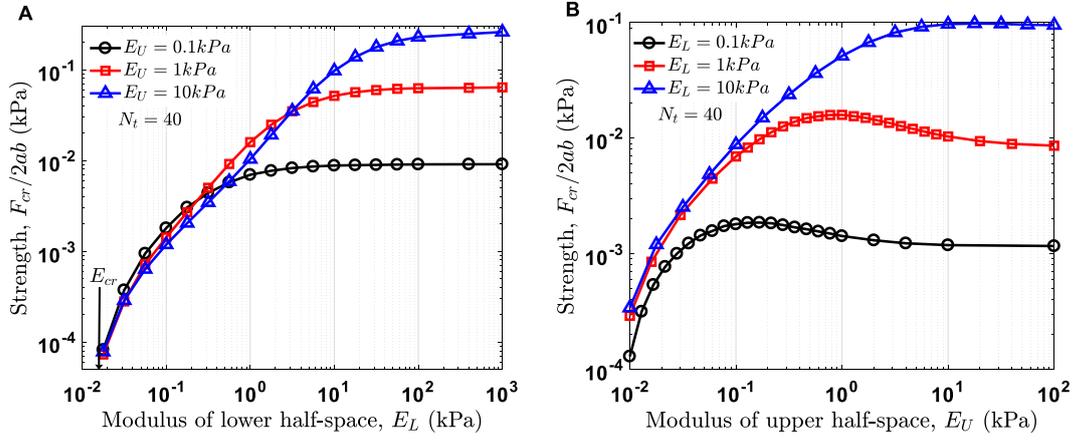

**Fig. 7.** Cluster strength as a function of the modulus of the lower and upper half-spaces. The total number of molecular bonds is $N_t = 40$. The black circle, red square, and blue triangle correspond to the upper/lower half-space with Young's modulus of 0.1, 1, and 10 kPa.

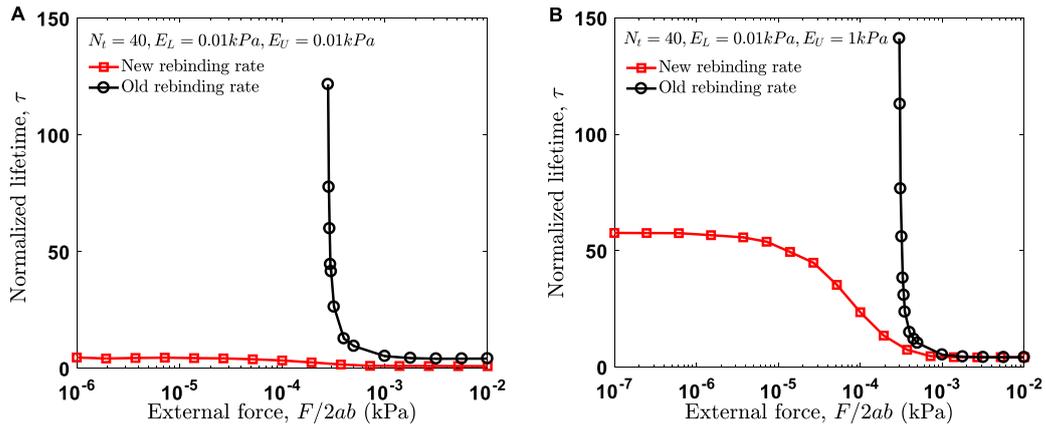

**Fig. 8.** Normalized lifetime of the cluster as a function of external force under the soft lower half-space ($E_L = 0.01$ kPa). The modulus of the upper half-space is 0.01 kPa in (*A*) and 1 kPa in (*B*). Black circle is plotted by the previous rebinding rate (Qian et al. [16], Eq. 29), and red square is plotted by the newly proposed rebinding rate (Eq. 27).

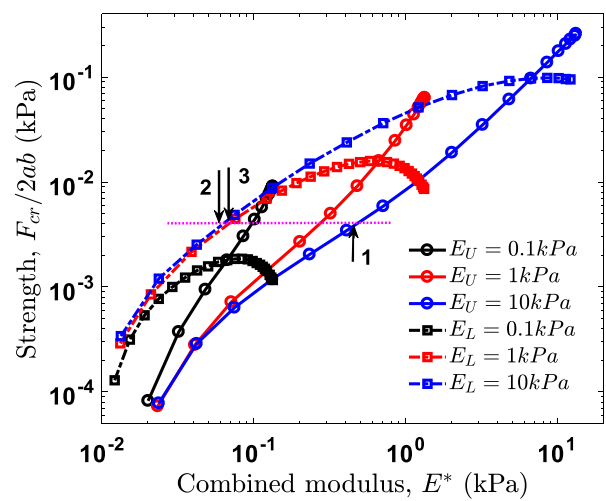

**Fig. 9.** Replot of Fig. 7 as a function of combined modulus $E^*$.

**Table 1.** List of parameters used in the text.

| Parameters | Values |
| --- | --- |
| focal adhesion size, $2a$ ($\mu$m) | 0.32~3.2 |
| spacing between neighboring bonds, $b$ (nm) | 32 |
| total number of bonds, $N_t = 2a/b$ | 2~40 |
| Poisson ratio of elastic half-spaces, $\nu_U, \nu_L$ | 0.5 |
| binding radius, $l_{bind}$ (nm) | 1 |
| single bond stiffness, $k_{LR}$ (pN/nm) | 0.25 |
| elastic modulus of upper half-space, $E_U$ (kPa) | $10^{-2}$~$10^2$ |
| elastic modulus of lower half-space, $E_L$ (kPa) | $10^{-2}$~$10^3$ |
| force scale in bond dissociation, $F_b$ (pN) | 4 |
| radius of individual bonds, $a_0$ (nm) | 5 |

| ratio of reaction rates, $k_{on}^0 / k_0$ | 1 |